\documentclass[9pt,twoside,twocolumn]{pnas-new}

\templatetype{pnasresearcharticle} 
\usepackage{graphicx,grffile}
\usepackage{amsmath,amssymb,siunitx,booktabs,multimedia,bm}
\usepackage{caption}
\usepackage{subcaption}
\usepackage{tikz}
\usetikzlibrary{arrows.meta,decorations.markings,patterns,shapes.geometric}
\usepackage{xcolor}
\usepackage{algorithm,algpseudocode}
\usepackage{physics}
\usepackage{hyperref}
\usepackage{threeparttable} 

\begin{document}
\title{Generative Monte Carlo Sampling for Constant-Cost Particle Transport}

\author[a]{Joseph A. Farmer}
\author[a]{Aidan Murray}
\author[a]{Johannes Krotz}
\author[a,1]{Ryan G. McClarren}

\affil[a]{Department of Aerospace and Mechanical Engineering, University of Notre Dame, Notre Dame, Indiana 46637, USA}

\leadauthor{Farmer}

\significancestatement{ 
\textcolor{black}{For nearly ten decades, Monte Carlo methods have served as the gold standard for particle transport in nuclear reactor design, cancer radiotherapy, astrophysical modeling, and inertial confinement fusion, yet their computational cost has remained tethered to the number of scattering events a particle undergoes. We introduce Generative Monte Carlo (GMC), which breaks this coupling. By training a neural network to sample particle exit states directly, GMC bypasses the random walk entirely, achieving constant cost per cell transmission regardless of optical thickness. A single trained model generalizes across all materials, and because the method reduces transport simulation to neural-network inference, it inherits every advance in AI hardware and algorithms. As accelerators grow faster and architectures more efficient, GMC simulations accelerate in lockstep without modifying the underlying physics code. This creates a future-proof pathway for scalable, high-fidelity modeling in complex materials. }
}

\authorcontributions{R.G.M.~designed research; J.A.F., A.M. and J.K. performed research; E.H. and L.Z. contributed new reagents/analytic tools; J.A.F., J.K., and R.G.M. analyzed data; and J.A.F., J.K., and R.G.M. wrote the paper. }
\authordeclaration{The authors declare no competing interests.}
\correspondingauthor{\textsuperscript{1}To whom correspondence should be addressed. E-mail: rmcclarr@nd.edu}

\keywords{Monte Carlo $|$ particle transport  $|$ machine learning $|$ generative artificial intelligence $|$ conditional flow matching}

\begin{abstract}
We present Generative Monte Carlo (GMC), a novel paradigm for particle transport simulation that integrates generative artificial intelligence directly into the stochastic solution of the linear Boltzmann equation. By reformulating the cell-transmission problem as a conditional generation task, we train neural networks using conditional flow matching to sample particle exit states, including position, direction, and path length, without simulating scattering histories. The method employs optical coordinate scaling, enabling \textit{a single trained model} to generalize across any material. We validate GMC on two canonical benchmarks, namely a heterogeneous lattice problem characteristic of nuclear reactor cores and a linearized hohlraum geometry representative of high-energy density radiative transfer. Results demonstrate that GMC preserves the statistical fidelity of standard Monte Carlo, exhibiting the expected $1/\sqrt{N}$ convergence rate while maintaining accurate scalar flux profiles. While standard Monte Carlo computational cost scales linearly with optical thickness in the diffusive limit, GMC achieves constant $O(1)$ cost per cell transmission, yielding order-of-magnitude speedups in optically thick regimes. This framework strategically aligns particle transport with modern computing architectures optimized for neural network inference, positioning transport codes to leverage ongoing advances in AI hardware and algorithms.
\end{abstract}

\dates{This manuscript was compiled on \today}

\maketitle
\thispagestyle{firststyle}

\section*{Introduction}

\dropcap{M}onte Carlo (MC) methods for particle transport have long been regarded as the highest-fidelity tools for modeling a variety of physical systems: neutron behavior in fission reactor systems, rarefied gas dynamics with Direct Simulation Monte Carlo \cite{bird1994molecular}, x-ray radiative transfer in high-energy-density experiments, and a range of other multiscale streaming-and-scattering phenomena~\cite{jacoboni2012monte,cahalan2005i3rc}. MC's computational demands have historically driven algorithms to evolve alongside hardware, from vectorized implementations on Cray supercomputers in the 1980s~\cite{brown1984monte} to modern event-based algorithms for heterogeneous CPU-GPU architectures \cite{pandya2016implementation}. Yet, despite extraordinary advances in hardware, the fundamental computational bottleneck of Monte Carlo---resolving every scattering interaction---remains unchanged.

MC simulations have also historically motivated investment in high-performance computing platforms. The first petascale computer, Roadrunner, counted MC radiative-transfer solvers among its flagship applications~\cite{gray2008roadrunner}, and earlier US federal programs such as the Accelerated Strategic Computing Initiative explicitly identified MC as a target workload~\cite{llnl1999asci}. Over time, however, the influence of scientific simulation on hardware design has diminished. In the 1990s and 2000s, government-funded supercomputing represented a substantial fraction of the overall market, allowing the needs of scientific computing to directly shape vendor roadmaps. As the commercial demand for high-throughput accelerators grew---particularly for graphics and later machine-learning workloads---scientific applications increasingly had to adapt to hardware not designed with them in mind. Today, the rapid expansion of artificial intelligence and machine learning (AI/ML) has shifted hardware innovation decisively toward architectures optimized for neural-network inference and training.

Recent advances in generative modeling offer a new opportunity to rethink particle-transport simulation at this more fundamental level. Normalizing flows~\cite{rezende2015variational,papamakarios2021normalizing} and diffusion-based models have transformed probabilistic modeling in machine learning by learning mappings from simple base distributions to complex target distributions, enabling efficient sampling and exact likelihood evaluation. These methods have demonstrated success across diverse areas of scientific computing, including molecular dynamics~\cite{noe2019boltzmann}, lattice field theory~\cite{kanwar2020equivariant}, and Bayesian inference in physical systems~\cite{cranmer2019frontier}. A recent development, Conditional Flow Matching (CFM)~\cite{lipman2022flow,liu2022flow}, simplifies the training of continuous normalizing flows by directly regressing velocity fields, avoiding the need for stochastic simulation during training and offering computational advantages over score-based diffusion models~\cite{song2020score}.

\textcolor{black}{Neural networks have recently been applied in several was to particle transport. One thrust has focused on operator learning, training neural networks to map directly from problem parameters to solution fields. Fourier Neural Operators \cite{li2021fourier} and related architectures have demonstrated accuracy improvements and speed-ups in high-energy density radiative transfer~\cite{farmer2024fourier} and nonlinear radiation diffusion~\cite{lin2024efficient}. A parallel line of work addresses inverse problems, using deep learning to retrieve temperature concentration and distributions from spectral measurements~\cite{ren2019machine} or to replace line-by-line transmittance calculations in atmospheric remote sensing~\cite{stegmann2022deep}. Generative models have also begun to appear in Monte Carlo workflows, where generative adversarial networks have been used to compress and resample posterior ensembles from Bayesian geophysical inversions~\cite{scheiter2022upscaling} and  to limit the degrees of freedom in Monte Carlo economic studies.  Despite this progress, no method has emerged that preserves the full statistical fidelity of Monte Carlo while accelerating simulation across materials and physical regimes.  }

Therefore, just as in the past, innovation is needed in MC algorithms to take full advantage of computing hardware without sacrificing physics fidelity. This paper presents such a new paradigm. Our novel approach is to use ML as a native MC sampler. Rather than following the random walk of particles, we train a neural network to produce samples from a conditional distribution that matches the distribution that would arise from many samples of a standard MC calculation. The ML models are based on generative AI methods and require only a single trained model for a given geometry class; for example, for two-dimensional neutron transport, one trained model can solve any problem within the training distribution. The specific model architecture is not essential, any ML architecture capable of producing samples from a target conditional distribution can be used in our approach given appropriate training data. To our knowledge, this work represents the first use of generative models as direct samplers within the Monte Carlo algorithm, enabling a single trained model to accelerate particle transport across a broad range of application areas where Monte Carlo methods are employed.

\textcolor{black} {Monte Carlo methods emerged from the challenge of understanding neutron behavior in fissionable materials in the 1930's. Enrico Fermi had experimented with statistical sampling and built the FERMIAC --- a Monte Carlo trolley --- to trace neutron genealogies mechanically~\cite{metropolis1987beginning}. The method took its modern form when Metropolis, Ulam, and von Neumann formalized it for electronic computers~\cite{metropolis1949monte}. In the decades since, practitioners have followed a naming convention, where Monte Carlo variants are distinguished by descriptive modifiers. Implicit Monte Carlo~\cite{fleck1971implicit} introduced effective scattering to stabilize radiative transfer. Quasi-Monte Carlo~\cite{niederreiter1992random,caflisch1998monte,farmer2020quasi} and Randomized Quasi-Monte Carlo~\cite{owen1995randomly,lecuyer2002recent,pasmann2025randomized} exploit low-discrepancy sequences for faster convergence. Sequential Monte Carlo,~\cite{doucet2001sequential} addressed filtering in dynamical systems, Multilevel Monte Carlo~\cite{giles2008multilevel,giles2015multilevel} introduced hierarchical variance reduction, and Markov Chain Monte Carlo~\cite{metropolis1953equation,hastings1970monte} enabled sampling from complex equilibrium distributions. Each variant introduced new efficiencies or capabilities while preserving the foundational principle of statistical estimation through random sampling. In this spirit, we refer to our generative-assisted method as Generative Monte Carlo (GMC), where neural networks learn conditional distribution governing particle transmission and sample particle exit states directly.}

An ML-based MC algorithm has two key consequences. First, it allows particle transport calculations to directly benefit from hardware and algorithmic advances driven by the explosion of investment in AI/ML. As hardware improves, so will our efficiency, and breakthroughs in ML algorithms will similarly benefit our approach. Second, the performance of software implementing MC algorithms will be primarily driven by the accuracy and efficiency of ML models --- the best models will yield the best simulations.

A sketch of the new MC algorithm for particle transport can be given in a few steps. Assume the system in question is described by the density of streaming and scattering particles. Consider a cell in a computational mesh inside which the material properties are constant. When a particle enters that cell, in the absence of absorption, the particle will eventually exit after some number of scattering events. Upon exiting, the particle will have path length $s$ in the cell and will exit with a particular direction $\mathbf{\Omega}$. Standard MC would simulate each of the scattering events inside the cell. In our approach, we use ML to directly sample $s$ and $\mathbf{\Omega}$ for the particle when it leaves the cell, conditioned on where the particle entered the cell and its original direction as well as the dimensions of the cell. We then exploit the property that a purely scattering solution can be scaled to a problem with absorption by simple exponential attenuation. As a result, we can follow particles from cell to cell without simulating their interactions inside each cell. Because the sampling cost is constant per particle, simulations take the same number of operations regardless of the optical thickness regime. This has obvious benefits in reduced computational cost and the ability to readily load balance parallel calculations.

Our approach is redolent of previous MC algorithms that combine many particle events into a single step to reduce computational cost. These include the condensed history method common in electron transport~\cite{kawrakow2000accurate} and the random walk method~\cite{fleck1984random} used in diffusive radiative transfer calculations. However, those earlier methods either sacrifice fidelity to achieve performance gains or are applicable only in particular physical regimes. The condensed history method relies on analytical approximations such as Moli\`ere theory that are valid only for specific scattering processes, while the random walk method exploits the diffusion approximation and loses accuracy in streaming-dominated regions. Our approach requires no such approximations, the neural network learns the exact conditional distribution from training data generated by standard Monte Carlo, preserving full physics fidelity across all transport regimes. The method is thus approximation-free in the sense that accuracy is limited only by the neural network's capacity to represent complex distributions---a limitation that can be systematically reduced by increasing model  and training data size.

\section*{Theoretical Background}

\subsection*{The Transport Equation}

The simulations in this work are governed by the steady-state, monoenergetic linear Boltzmann transport equation. This integro-differential equation enforces particle conservation within an infinitesimally small volume of phase space defined by position $\mathbf{r}$ and direction $\mathbf{\Omega}$. For a spatial domain $D$ with boundary $\partial D$, the angular flux $\psi(\mathbf{r},\mathbf{\Omega})$ satisfies
\begin{multline}
\mathbf{\Omega}\cdot\nabla\psi(\mathbf{r},\mathbf{\Omega})
+ \sigma_t(\mathbf{r})\psi(\mathbf{r},\mathbf{\Omega}) = \\
\int_{\mathbb{S}^2} \sigma_s(\mathbf{r},\mathbf{\Omega}' \cdot \mathbf{\Omega})
\psi(\mathbf{r},\mathbf{\Omega}') \, d\mathbf{\Omega}'
+ Q(\mathbf{r},\mathbf{\Omega}),
\end{multline}
where material properties are characterized by the total cross section $\sigma_t$ and the scattering cross section $\sigma_s$. The absorption cross section is defined as $\sigma_a = \sigma_t - \sigma_s$. The equation states that the net loss of particles from a phase-space volume due to streaming and collisions (left-hand side) must be balanced by the gain of particles scattering into direction $\mathbf{\Omega}$ from all other directions $\mathbf{\Omega}'$, plus any independent particle production $Q$.

While the angular flux contains the complete description of the radiation field, the quantity of primary physical interest in many applications is the scalar flux, $\phi(\mathbf{r})$, defined as the integral of the angular flux over the unit sphere
\begin{equation}
\phi(\mathbf{r}) = \int_{\mathbb{S}^2} \psi(\mathbf{r}, \mathbf{\Omega})\, d\mathbf{\Omega}.
\end{equation}
Physically, the scalar flux represents the total path length traveled by all particles per unit volume per unit time. Its importance stems from the fact that most interaction probabilities in bulk media---such as absorption or fission---are independent of the incident particle's direction. Consequently, the reaction rate density $R_x(\mathbf{r})$ for a process with macroscopic cross section $\sigma_x$ is determined by the product of the cross section and the scalar flux
\begin{equation}
R_x(\mathbf{r}) = \sigma_x(\mathbf{r}) \, \phi(\mathbf{r}).
\end{equation}
The scalar flux thus links the kinetic transport solution to engineering quantities of interest. In nuclear reactor physics, the local power density is proportional to the fission rate, which is calculated using the scalar flux because fuel nuclei absorb neutrons isotropically. In medical physics, the scalar flux is essential for determining the absorbed dose rate, as the aggregate energy deposited in tissue depends on the total fluence of ionizing particles rather than their individual trajectories. Likewise, in high-energy density physics and radiative heat transfer, the scalar flux is required to calculate the heating rate due to absorption of radiation.

\subsection*{Monte Carlo Method}

The Monte Carlo (MC) method provides a numerical solution to the transport equation by constructing a stochastic random walk that satisfies the integral form of the Boltzmann equation. Unlike deterministic methods that solve for the average flux directly, MC simulates a large number of individual particle histories and aggregates their contributions to estimate ensemble averages. This approach reconstructs the macroscopic solution by accumulating contributions from discrete particles as they undergo varying numbers of scattering events throughout the system.

The simulation initializes by sampling a particle's position and direction from the source distribution $Q(\mathbf{r}, \mathbf{\Omega})$. Each simulated particle is assigned an initial energy weight $E$ representing the total number of physical particles carried by the computational particle packet. To maximize computational efficiency, this work employs a continuous absorption scheme, often termed implicit capture. Unlike analog MC, which terminates particles discretely upon absorption, this method treats absorption analytically along the particle's trajectory, allowing the particle to explore phase space for longer durations.

Particle streaming is modeled via ray tracing. A particle starting at position $\mathbf{r}$ inside a given simulation cell with direction $\mathbf{\Omega}$ moves along this direction until it either undergoes a scattering event or leaves the cell. The distance to the geometric boundary is denoted $d_b$. Because absorption is handled continuously, the stochastic sampling of collision sites is determined by the scattering cross section $\sigma_s$. The distance to the next scattering event, $d_s$, is sampled from an exponential distribution
\begin{equation}
d_s = -\frac{\ln(\xi)}{\sigma_s},
\end{equation}
where $\xi \in (0,1]$ is a uniform random variate. The particle's track length for the current step is $\Delta s = \min(d_b, d_s)$. If $\Delta s = d_s$, the particle undergoes a scattering event in which its direction is updated according to the scattering kernel; otherwise, it streams to the boundary and enters the adjacent cell.

As the particle streams through a medium with absorption cross section $\sigma_a$, the weight $w$ is attenuated continuously according to the Beer-Lambert law. Suppose a particle traverses a cell in $N$ steps with track lengths $\Delta s_1, \ldots, \Delta s_N$ between events, yielding a total track length $\Delta s_{\text{tot}} = \sum_{i=1}^N \Delta s_i$. If the particle enters a cell with  weight $w_{\text{in}}$, its  weight upon exiting the cell is
\begin{equation}
w_{\text{out}} = w_{\text{in}}\, e^{-\sigma_a \Delta s_{\text{tot}}}.
\end{equation}
Particle conservation then gives that the absorption in the material medium is
\begin{equation}
\Delta w = w_{\text{in}} \left( 1 - e^{-\sigma_a \Delta s_{\text{tot}}} \right).
\end{equation}

The scalar flux is estimated by tallying the deposited weight per unit volume and normalizing by the absorption cross section. For a mesh cell of volume $V$, the contribution to the scalar flux from this particle's track segment is
\begin{equation}
\phi_{\text{est}} = \frac{\Delta w}{V \sigma_a}.
\end{equation}
This estimator is statistically robust, and the final scalar flux solution is obtained by accumulating these contributions over all tracks of all simulated histories. A critical observation motivating our approach is that while the standard MC solver must simulate complete trajectories with potentially many scattering events, the flux estimator $\phi_{\text{est}}$ depends exclusively on the total track length $\Delta s_{\text{tot}}$. The individual scattering events affect only the distribution of $\Delta s_{\text{tot}}$ and the exit state, not the functional form of the estimator. This observation suggests that if one could directly sample the total track length and exit state from the correct conditional distribution, the intermediate scattering events need not be simulated explicitly.

\section*{Generative Monte Carlo Sampler}

\subsection*{Framework}

Generative Monte Carlo (GMC) accelerates particle transport by replacing the stochastic sampling of collisions with a learned macroscopic mapping. Rather than generating detailed particle trajectories, we learn to sample the total track length directly. While standard Monte Carlo integrates the transport equation by simulating every scattering event within a mesh element, GMC treats cell transmission as a generative modeling task. The core engine is Conditional Flow Matching (CFM), which learns a continuous time-dependent velocity field that transforms Gaussian noise into the conditional distribution of particle exit states.

\subsubsection*{Conditional Flow Matching}

Formally, we aim to sample from the conditional probability distribution $p(\mathbf{y} | \mathbf{c})$, where $\mathbf{c}$ represents the particle entry state and cell geometry, and $\mathbf{y}$ represents the exit state. CFM achieves this by learning a velocity field $v_\theta(\mathbf{x}, t)$ for $t \in [0,1]$. The inference process generates a sample by solving the ordinary differential equation (ODE)
\begin{equation}
\frac{d\mathbf{x}}{dt} = v_\theta(\mathbf{x}, t), \quad \mathbf{x}(0) \sim \mathcal{N}(0, \mathbf{I}).
\end{equation}
Integrating this field from $t=0$ to $t=1$ yields a sample $\mathbf{x}(1)$ distributed according to the learned physics of particle transport $p(\mathbf{y} | \mathbf{c})$.

The velocity field is trained using the flow matching objective, which minimizes the mean squared error between the predicted velocity and the optimal transport velocity along interpolated paths. Given conditioning variables $\mathbf{c}$ and target exit states $\mathbf{y}$, the loss function takes the form
\begin{equation}
\mathcal{L}(\theta) = \mathbb{E}_{t \sim \mathcal{U}[0,1], \mathbf{x}_0, \mathbf{x}_1} \left[ \| v_\theta(\mathbf{x}_t, t) - \dot{\mathbf{x}}_t \|^2 \right],
\end{equation}
where $\mathbf{x}_t = (1-t)\mathbf{x}_0 + t\mathbf{x}_1$ interpolates between noise samples $\mathbf{x}_0$ and target data $\mathbf{x}_1$ under a conditional optimal transport scheduler. This formulation enables efficient training without requiring simulation of the full ODE during the backward pass, a key advantage over earlier continuous normalizing flow approaches.

While we employ CFM in our implementation, other generative modeling frameworks capable of sampling from conditional distributions could serve the same role. The GMC framework is agnostic to the specific generative model employed; the essential requirement is faithful reproduction of the conditional exit distribution $p(\mathbf{y} | \mathbf{c})$.

\subsubsection*{Network Architecture and Inputs}

To handle the high dimensionality of phase space and varying material properties, we utilize two distinct neural networks. An \texttt{Internal Model} processes particles born from volumetric sources within mesh cells, while a \texttt{Boundary Model} handles particles streaming between adjacent cells. Based on the underlying physics, we parameterize the conditioning vector $\mathbf{c}$ using optical coordinates to ensure the model \textit{generalizes across materials without retraining}. For a rectangular cell of physical width $W$ and height $H$ with scattering cross section $\sigma_s$, the condition vector is defined as
\begin{equation}
\mathbf{c} = (\tilde{W}, \tilde{H}, \bm{\xi}_{\text{in}}, \mathbf{\Omega}_{\text{in}}),
\end{equation}
where $\tilde{W} = W \sigma_s$ and $\tilde{H} = H \sigma_s$ are the optical dimensions in units of mean free paths. The coordinate $\bm{\xi}_{\text{in}}$ is the normalized position along the entry face for the boundary model or the two-dimensional internal position for the source model. The angular direction $\mathbf{\Omega}_{\text{in}}$ is encoded via polar coordinates to maintain the unit disk constraint on direction cosines. This optical parameterization is essential because expressing geometry in terms of mean free paths rather than physical lengths allows a single trained model to be applied to cells with different physical dimensions and cross sections, provided the optical thicknesses fall within the training distribution.

The network predicts a target vector $\mathbf{y}$ representing the particle's state upon exiting the cell,
\begin{equation}
\mathbf{y} = (p_{\text{exit}}, \mathbf{\Omega}_{\text{exit}}, s),
\end{equation}
where $p_{\text{exit}}$ is a perimeter coordinate that maps uniquely to both the exit face and position along that face, 
$\mathbf{\Omega}_{\text{exit}}$ defines the exit direction in polar coordinates, and $s$ is the path length in the cell. The use of an unwrapped perimeter coordinate for the exit position simplifies the learning task by matching the dimension of the cell surface. 

The velocity field $v_\theta(\mathbf{x}, t)$ is parameterized by a feedforward neural network that processes both the flow time $t$ and the conditioning vector $\mathbf{c}$ through learned embeddings. The architecture employs residual blocks with adaptive modulation conditioned on the physical context, allowing the network to adjust its internal representations based on the optical properties of the cell. Detailed specifications of the network architecture, including layer configurations, activation functions, and normalization schemes are provided in Supplementary Material S.1.

Training data is generated from standard Monte Carlo simulations of pure-scattering transport. For each sample, cell optical dimensions are drawn from a distribution covering expected problem regimes, entry states are sampled according to the appropriate boundary or source distributions, and a standard Monte Carlo random walk is executed until the particle exits the cell. The resulting exit state, comprising position, direction, and path length, is recorded as the training target. In our results, a dataset of $10^6$ particle histories was generated and split for training and validation. All training is performed on GPU hardware, with wall-clock times on the order of several hours for the configurations reported here. However, this training cost is amortized across all subsequent simulations. Once trained, a single model serves any problem whose local cell properties fall within the trained optical thickness range, eliminating the need for problem-specific training. Further information on the training procedure is in Supplementary Material S.2.

\subsection*{Transport Algorithm}

The GMC transport algorithm replaces the standard collision-by-collision random walk with a cell-to-cell neural inference loop, proceeding through four stages.

In the first stage, the GMC boundary model is trained on a canonical reference frame where particles always enter from the left face. This design choice minimizes model complexity by eliminating the need to learn separate representations for each entry face. When a particle crosses a boundary in the global mesh, its state is rotated into this canonical local frame before inference.

The second stage performs the neural inference. A random Gaussian noise vector $\mathbf{z} \sim \mathcal{N}(0, \mathbf{I})$ is sampled and concatenated with the local condition $\mathbf{c}$. The trained GMC model then integrates the learned velocity field via the ordinary differential equation $d\mathbf{x}/dt = v_\theta(\mathbf{x}, t)$ from $t=0$ to $t=1$ to predict the exit state $\mathbf{y}$. This single inference step accounts for all underlying scattering events, effectively teleporting the particle to the cell boundary without simulating intermediate collisions. The details of the inference procedure can be found in S.3 in the Supplementary Material.

The third stage applies absorption and accumulates flux tallies. Consistent with the theoretical framework described earlier, absorption is treated analytically rather than stochastically. Using the sampled path length $\Delta s$ and the local absorption cross section $\sigma_a$, the particle's  weight $w$ is attenuated by the factor $\exp(-\sigma_a \Delta s)$. The scalar flux contribution is then computed from the energy deposition $\Delta w = w_{\mathrm{in}}(1 - e^{-\sigma_a \Delta s})$ and accumulated into the appropriate mesh cell tally.

The fourth stage updates the particle state in global coordinates. The predicted exit position, decoded from the perimeter coordinate $p_{\mathrm{exit}}$, and the exit direction are rotated from the canonical frame back to the global coordinate system. The particle is advanced to the adjacent cell, and the process repeats. The simulation continues until the particle exits the global domain or its energy weight falls below a prescribed cutoff threshold. We note that this procedure would be the same for any geometry (e.g., 1-D spherical, 2-D cylindrical, or 3-D), only the trained model would be different for each of these geometries.

\begin{figure*}[htbp]
    \centering
    \begin{subfigure}[b]{0.68\textwidth}
        \centering
        \includegraphics[width=0.99\textwidth]{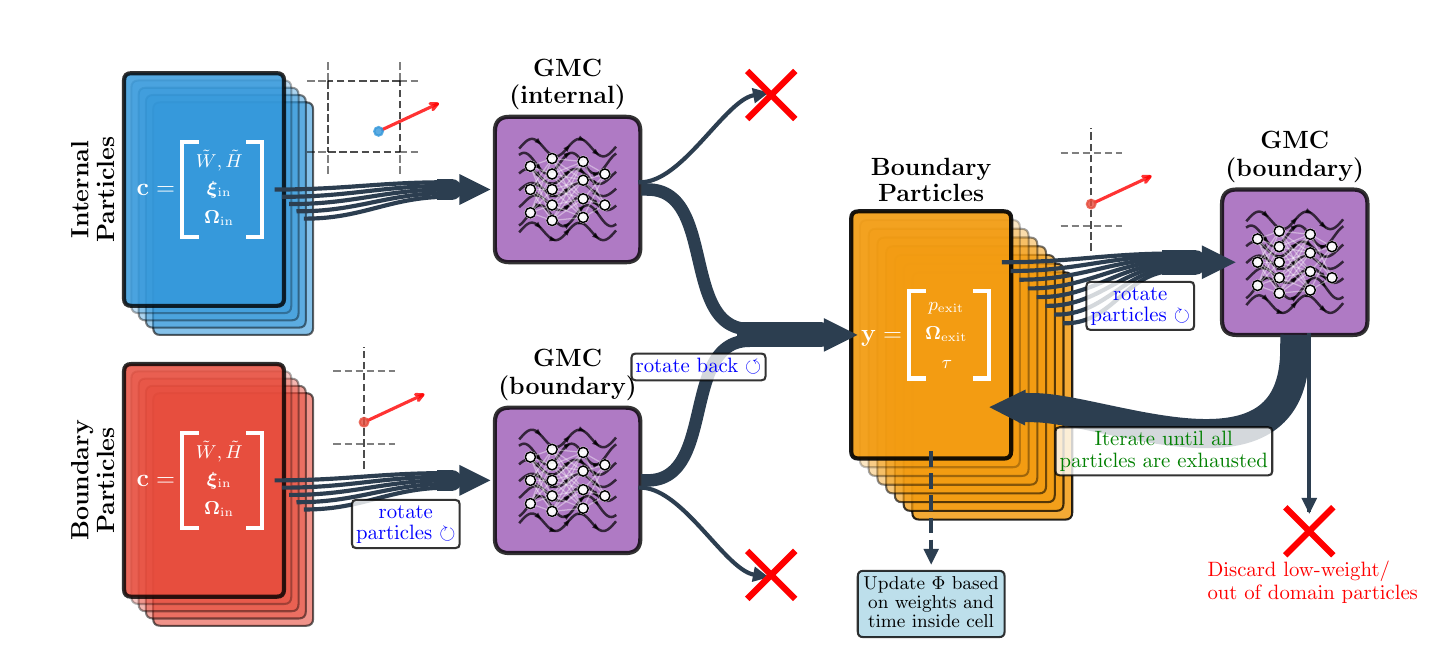}
        \caption{GMC transport workflow.}
        \label{fig:nf_workflow}
    \end{subfigure}
    \hfill
    \begin{subfigure}[b]{0.28\textwidth}
        \centering
        \includegraphics[width=0.99\textwidth]{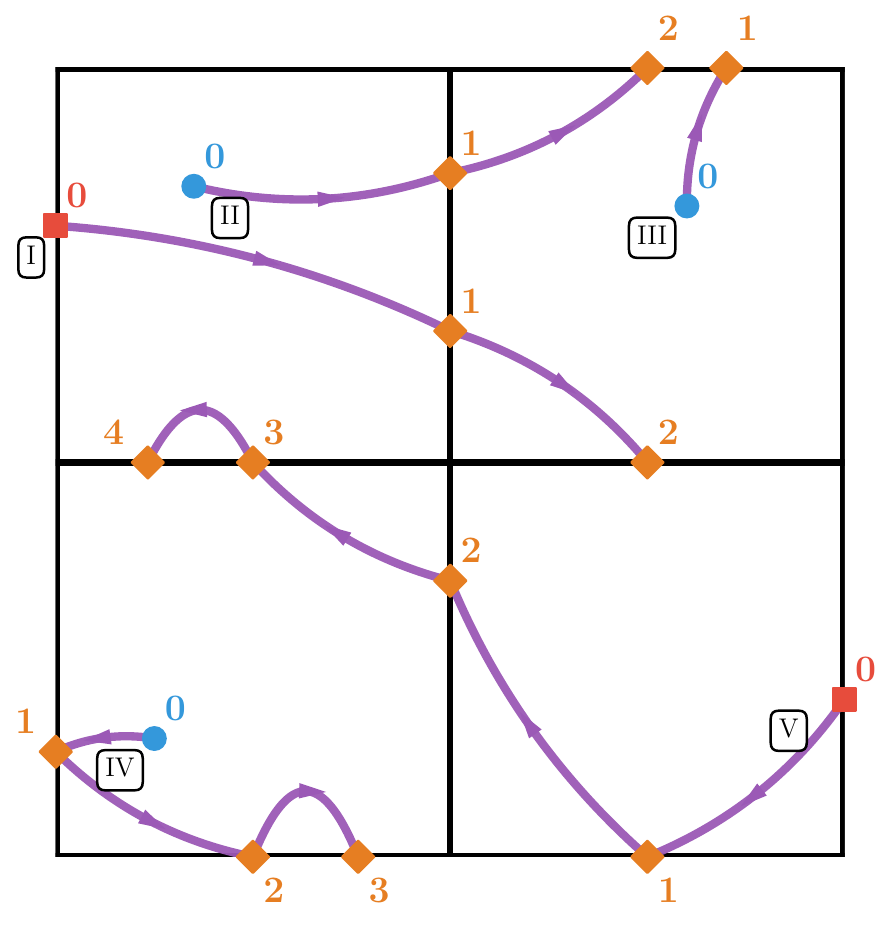}
        \caption{Flow trajectories across a mesh.}
        \label{fig:cfm_mc_2x2}
    \end{subfigure}

      \caption{%
    GMC workflow and representative trajectories.
    Figure~\ref{fig:nf_workflow} illustrates the transport pipeline. The conditioning vector $\mathbf{c} = (\tilde{W}, \tilde{H}, \bm{\xi}_{\text{in}}, \mathbf{\Omega}_{\text{in}})$ encodes optical dimensions, entry position, and direction; the output $\mathbf{y} = (p_{\mathrm{exit}}, \mathbf{\Omega}_{\text{exit}}, s)$ encodes exit position, direction, and path length.
    The workflow proceeds through four stages:
    (1)~Internal (blue) and boundary (red) particles are initialized.
    (2)~Separate networks process each type; boundary particles are rotated into the canonical left-entry frame.
    (3)~Particles are mapped to global coordinates and energy-weighted; those below threshold are terminated.
    (4)~The boundary model iterates on surviving particles until all exit the domain, accumulating scalar flux contributions.
    Figure~\ref{fig:cfm_mc_2x2} shows representative learned flow trajectories across a $2\times2$ mesh with colors consistent with the boxes in (a): Purple curves trace ODE integral paths from particles born inside a cell at $t=0$ (blue circles) or entering a cell face (red squares) and orange diamonds mark cell-to-cell transitions.}
    \label{fig:workflow_and_paths}
\end{figure*}

Figure~\ref{fig:workflow_and_paths} summarizes this pipeline. Figure~\ref{fig:nf_workflow} depicts the separation of internal source handling, which processes particles born from volumetric sources within mesh cells, and boundary-to-boundary transport, which governs particles streaming between adjacent cells. The internal model and boundary model are implemented as independent neural networks, each trained on data appropriate to its respective task. Figure~\ref{fig:cfm_mc_2x2} visualizes the resulting trajectories in a $2\times2$ mesh. The smooth integral curves represent the neural network's deterministic transformation in latent probability space; these curves correspond physically to complex, many-scatter random walks that the model has learned to bypass through direct conditional sampling.

\begin{figure*}[t]
  \centering
  \begin{subfigure}[b]{0.58\textwidth}
    \centering
    \includegraphics[width=0.95\textwidth]{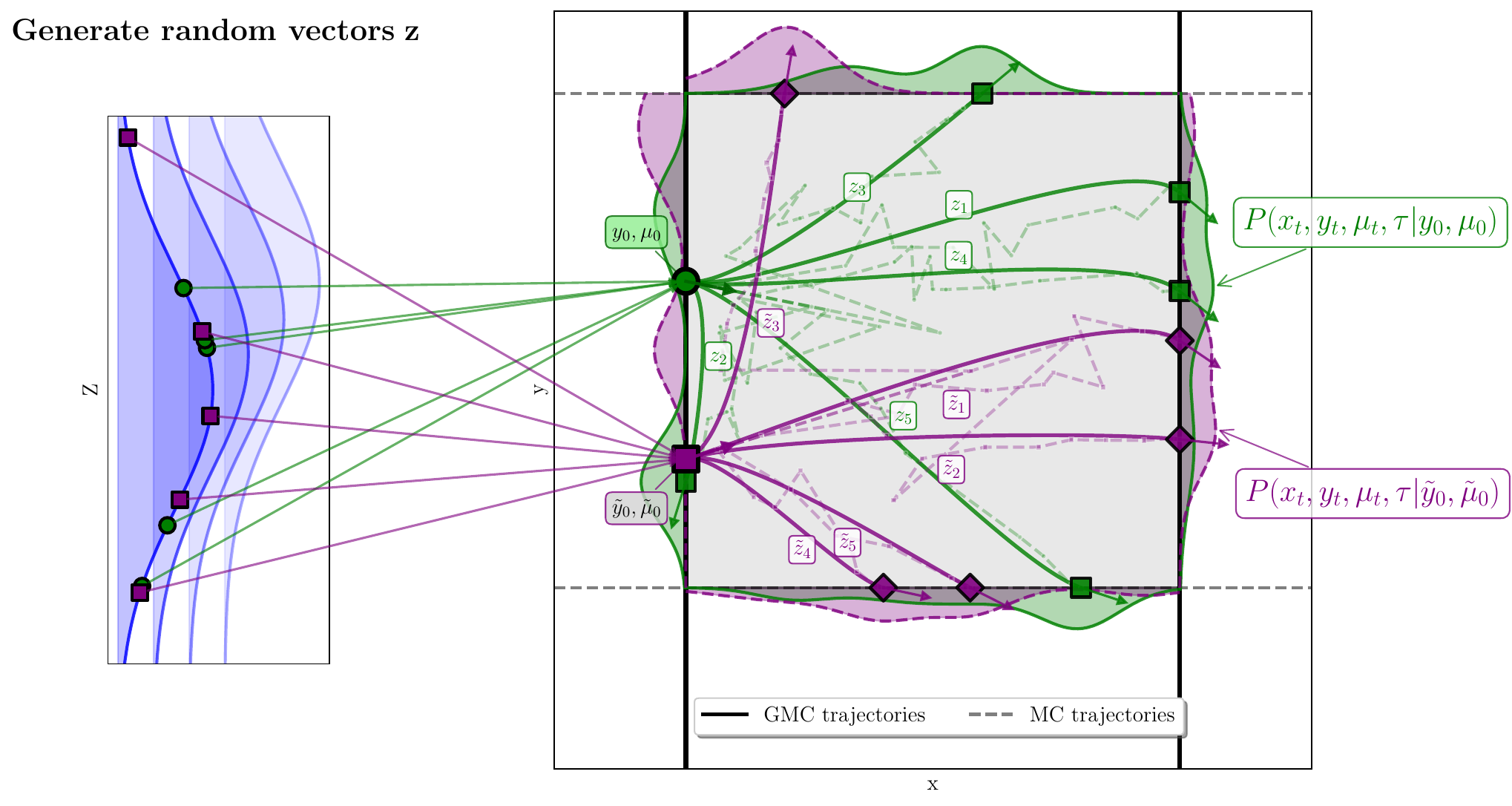}
    \caption{Conditional flow evolution in a slab.}
    \label{fig:slab_flow_matching}
  \end{subfigure}
  \hfill
  \begin{subfigure}[b]{0.38\textwidth}
    \centering
    \includegraphics[width=0.9\textwidth]{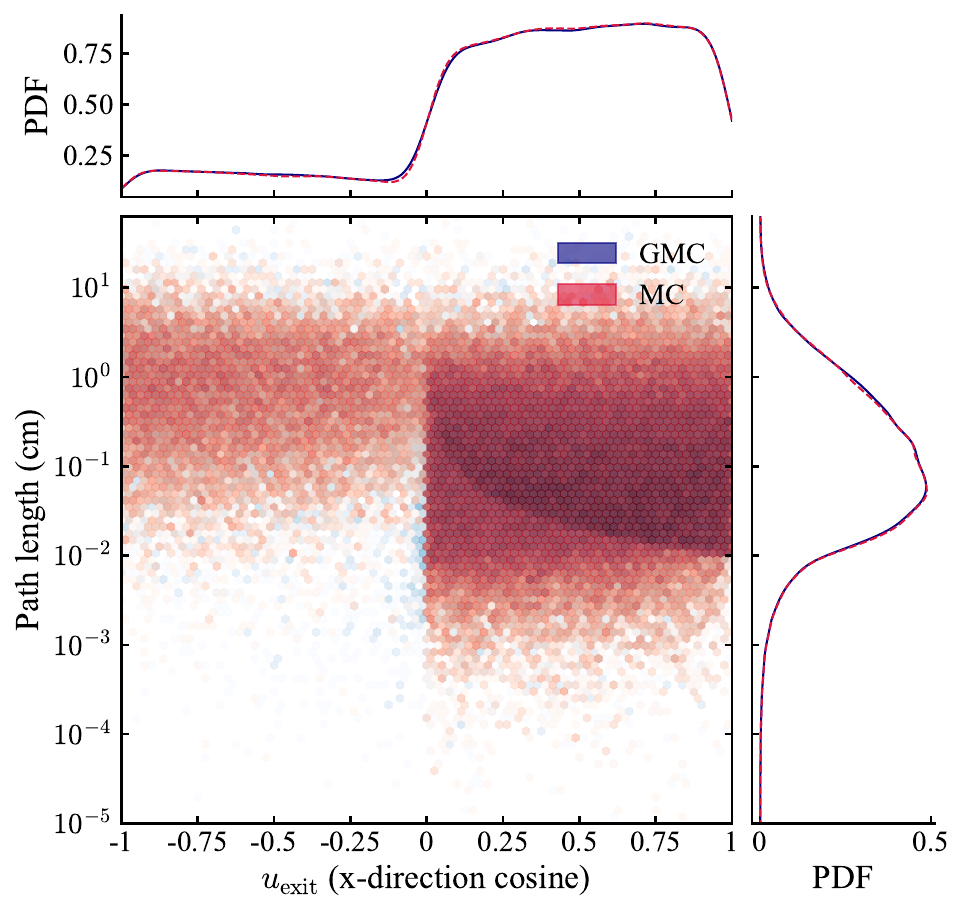}
    \caption{Comparison of GMC and MC statistics.}
    \label{fig:hexbin_mu_vs_logtime}
  \end{subfigure}

  \caption{%
  Representative conditional sampling behaviors.
  Figure~\ref{fig:slab_flow_matching} visualizes conditional flow integration in a homogeneous two-dimensional slab using the learned velocity field $v_\theta$.
  On the left, latent samples $z_i$ are drawn from a standard Gaussian distribution $\mathcal{N}(0, \mathbf{I})$.
  On the right, two distinct entry conditions are shown: $(y_0, \mu_0)$ in green and $(\tilde{y}_0, \tilde{\mu}_0)$ in purple, representing particles entering the cell at different positions and angles.
  For each condition, integrating $v_\theta$ from $t=0$ to $t=1$ transforms the latent samples into valid exit states.
  The smooth solid curves represent flow trajectories in the learned probability space; the dashed jagged lines depict a potential physical Monte Carlo random walk that the model bypasses.
  The marginal exit probability distributions along each boundary illustrate that both representations yield statistically equivalent ensemble behavior.
  Figure~\ref{fig:hexbin_mu_vs_logtime} compares the joint distribution of exit angle $u_{\mathrm{exit}}$ (x-direction cosine) versus $\log_{10}(\Delta s)$ (logarithmic path length) for standard MC (red density) and GMC (blue density), each generated with 80{,}000 samples.
  The hexbin density plot with marginals on the top and right axes demonstrates that GMC captures the complex correlations between exit angle and path length inherent in the transport physics.}
  \label{fig:conditional_sampling_panels}
\end{figure*}

To illustrate how the GMC framework functions as a transport surrogate, Figure~\ref{fig:slab_flow_matching} presents a schematic of the generative process. The left panel shows latent vector seeds $z_i$ sampled from the standard Gaussian base distribution $\mathcal{N}(0, \mathbf{I})$, which serve as the stochastic seeds for sample generation. Each latent vector will be transformed into a physically meaningful exit state through the learned flow.

The right panel illustrates the transformation of these seeds into physical transport trajectories within a two-dimensional Cartesian geometry. Two distinct entry conditions are depicted: $(y_0, \mu_0)$ in green and $(\tilde{y}_0, \tilde{\mu}_0)$ in purple, representing particles entering at different positions along the left boundary with different angular directions. The arrows connecting the latent distribution to the spatial domain signify the coupling of each random seed $z$ with the deterministic conditioning information to initialize the flow integration.

The trajectories in this figure highlight the fundamental difference between the surrogate model and analog Monte Carlo transport. The solid curves trace the integral curves of the learned velocity field $v_\theta(\mathbf{x}, t)$ as the flow parameter evolves from $t=0$ to $t=1$. These paths are smooth and, for a given seed, entirely deterministic; they represent the optimal transport of probability mass from the entry distribution to the exit distribution in latent space. In contrast, the dashed jagged lines depict the corresponding physical Monte Carlo histories that would be required in a standard  simulation. These paths reflect the stochastic zigzag trajectory arising from individual scattering events.

While the path of a single flow sample does not physically resemble the random walk of a particle undergoing multiple scattering events, the ensemble behavior of the two representations are statistically equivalent. The marginal exit probability distributions, shown along each slab boundary in the figure, show this similarity. Both the GMC flow and standard Monte Carlo sampling produce similar multi-modal exit distributions when aggregated over many samples. This statistical similarity ensures that the scalar flux, computed from energy deposition along particle tracks, is preserved across the computational domain. We present a deeper analysis of the generated distributions and potential bias in the sampling the in the Supplementary Material sections S.4 and S.6, respectively.

Figure~\ref{fig:hexbin_mu_vs_logtime} provides a direct comparison of the learned conditional distribution to that obtained from standard MC. This hexbin density plot compares the joint distribution of exit angle $u_{\mathrm{exit}}$, defined as the x-direction cosine of the exit direction, versus $\log_{10}(\Delta s)$, the logarithm of the path length, for single-cell transmissions. The marginal probability density functions are displayed on the top axis for the angular distribution and on the right axis for the path length distribution. The strong overlap between the Monte Carlo ground truth (shown in red) and the GMC predictions (shown in blue) across both the joint distribution and the marginal distributions confirms that the surrogate model faithfully captures the correlations between exit direction and path length. These correlations arise from the underlying transport physics where particles that scatter many times tend to have longer path lengths and emerge with a more isotropic angular distributions, while particles that scatter few times exit quickly with directions closer to their entry angles. The ability of GMC to reproduce these correlations is essential for the statistical fidelity required of a Monte Carlo estimator. Further details on the statistical fidelity of the distributions is contain in the Supplementary Material.

\section*{Results}

To demonstrate the GMC method's capability for realistic transport problems we compare solutions for a heterogeneous lattice characteristic of nuclear reactor cores, and a linearized hohlraum characteristic of high-energy density radiative transfer. Results comparing  GMC  against standard MC are presented in Figure~\ref{fig:validation}. For both benchmarks, the scalar flux $\phi(\mathbf{r})$ is computed using the track-length estimator described above. In addition to the results below, we present an algorithm and results for eigenvalue problems in Supplementary Material S.5. 

\subsection*{Benchmark Performance}

\begin{figure*}[p]
    \centering
    \newlength{\validationheight}
    \setlength{\validationheight}{5.5cm}

    \begin{subfigure}[b]{0.32\textwidth}
        \centering
        \includegraphics[height=\validationheight]{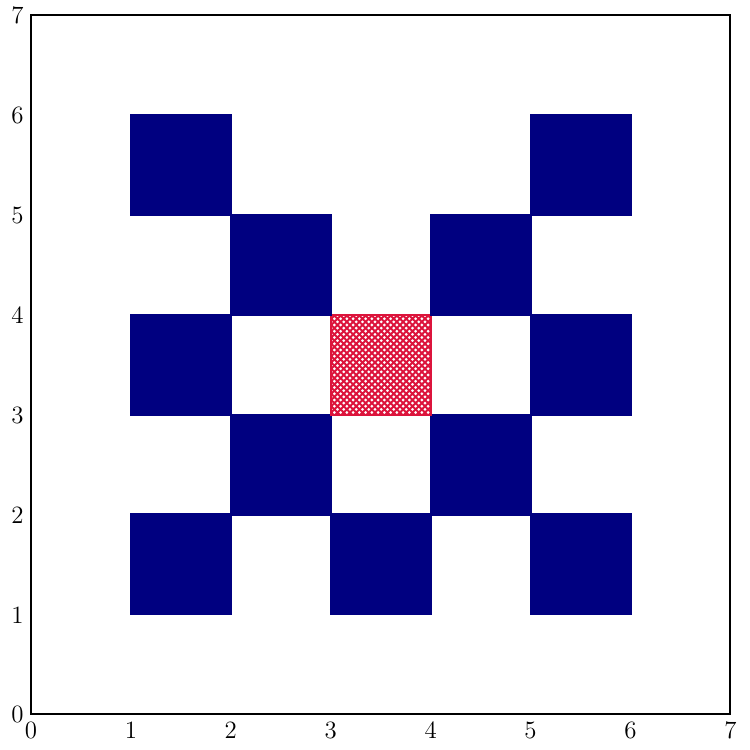}
        \caption{Geometry (distances in cm)}
    \end{subfigure}
    \hfill
    \begin{subfigure}[b]{0.32\textwidth}
        \centering
        \includegraphics[height=\validationheight]{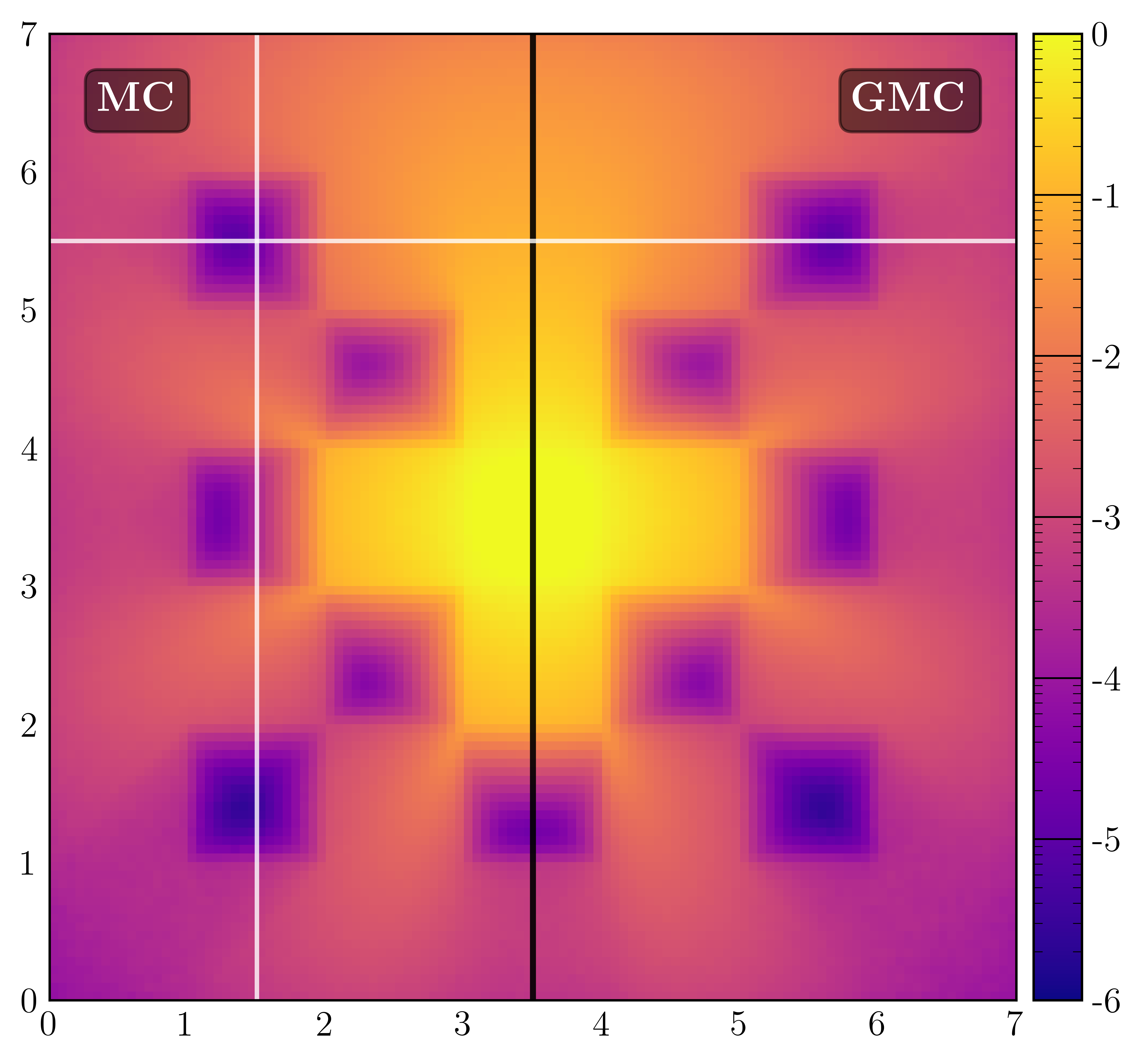}
        \caption{$\log_{10}$(Scalar flux)}
    \end{subfigure}
    \hfill
    \begin{subfigure}[b]{0.32\textwidth}
        \centering
        \includegraphics[height=\validationheight]{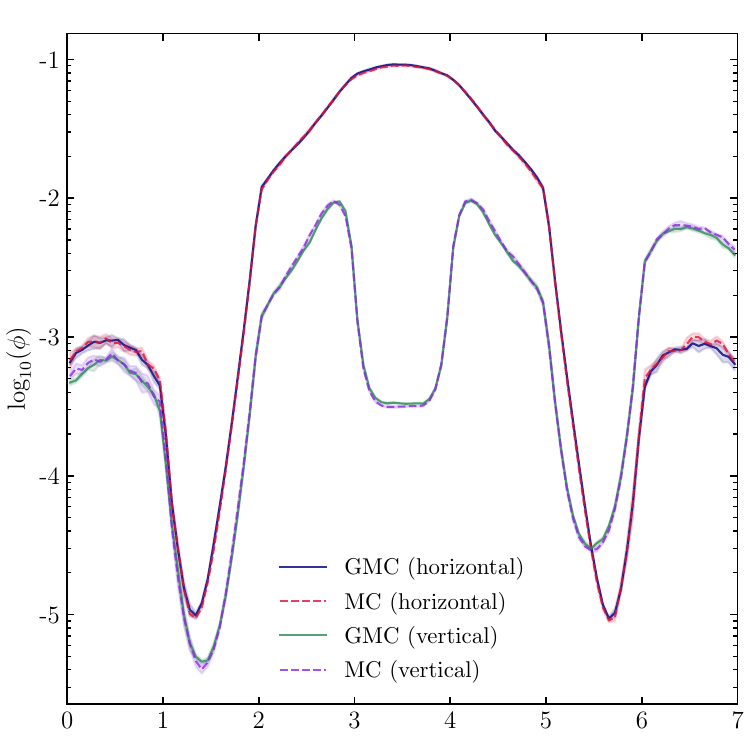}
        \caption{Lineout comparison}
    \end{subfigure}

    \vspace{0.5cm}

    \begin{subfigure}[b]{0.32\textwidth}
        \centering
        \includegraphics[height=\validationheight]{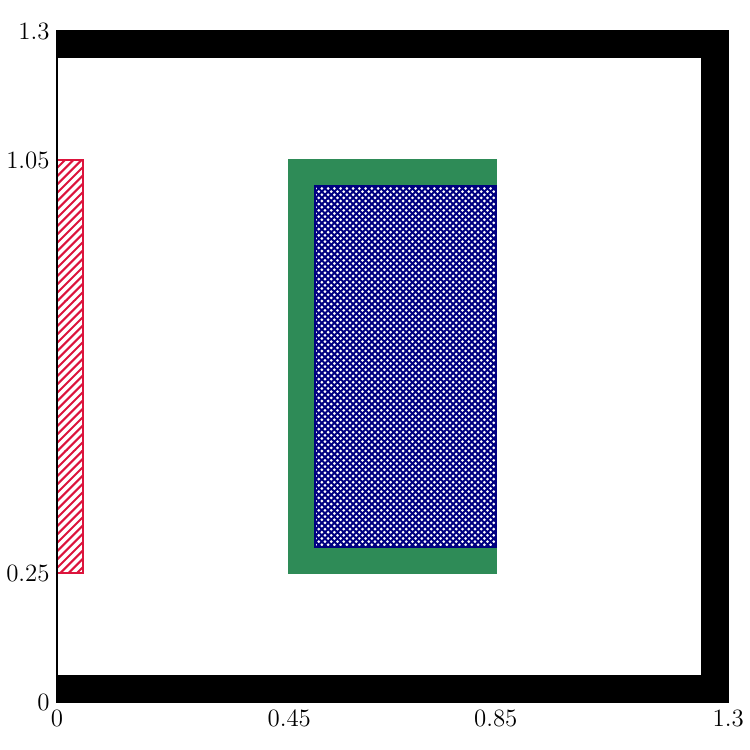}
        \caption{ Geometry (distances in cm)}
    \end{subfigure}
    \hfill
    \begin{subfigure}[b]{0.32\textwidth}
        \centering
        \includegraphics[height=\validationheight]{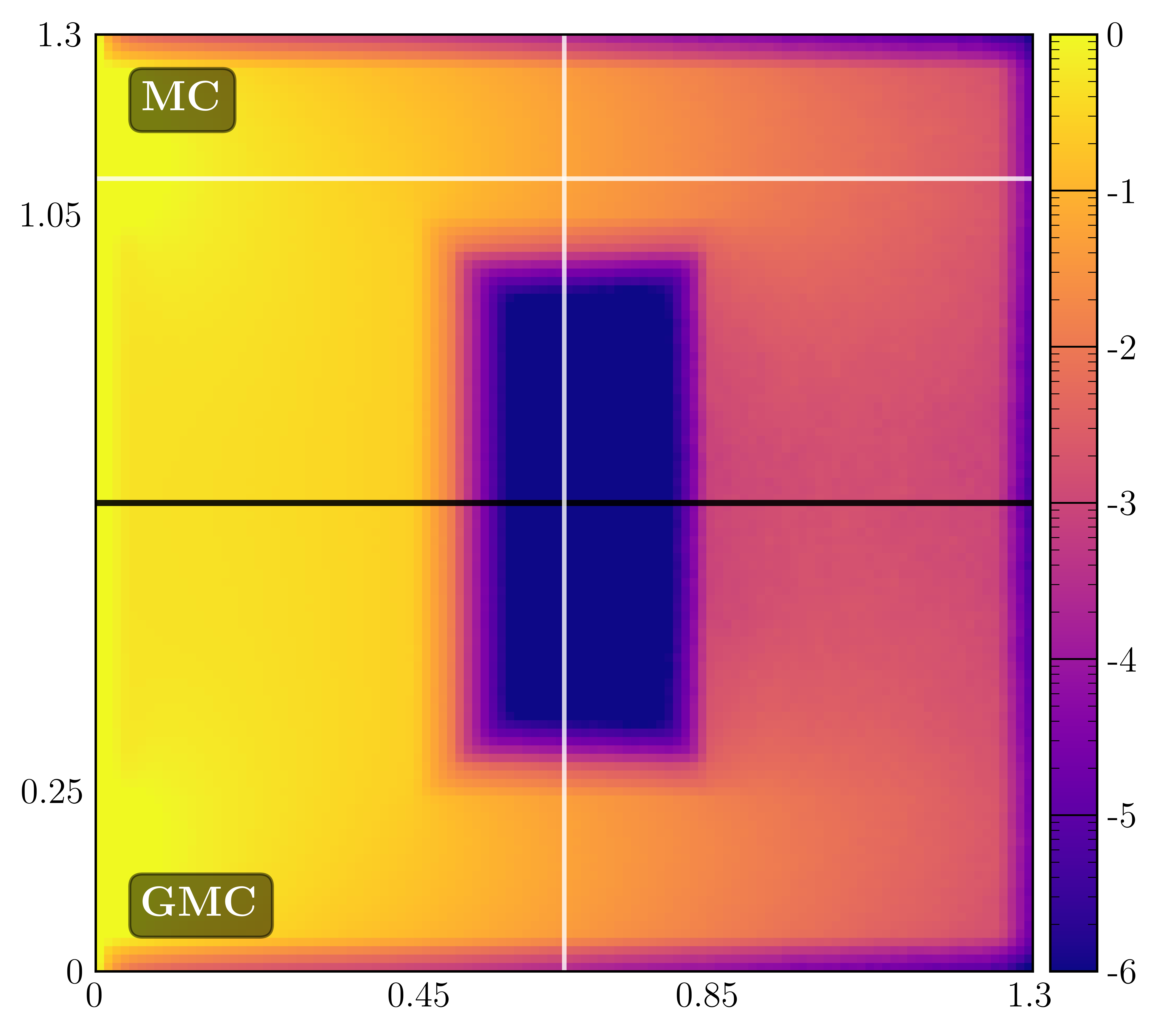}
        \caption{$\log_{10}$(Scalar flux)}
    \end{subfigure}
    \hfill
    \begin{subfigure}[b]{0.32\textwidth}
        \centering
        \includegraphics[height=\validationheight]{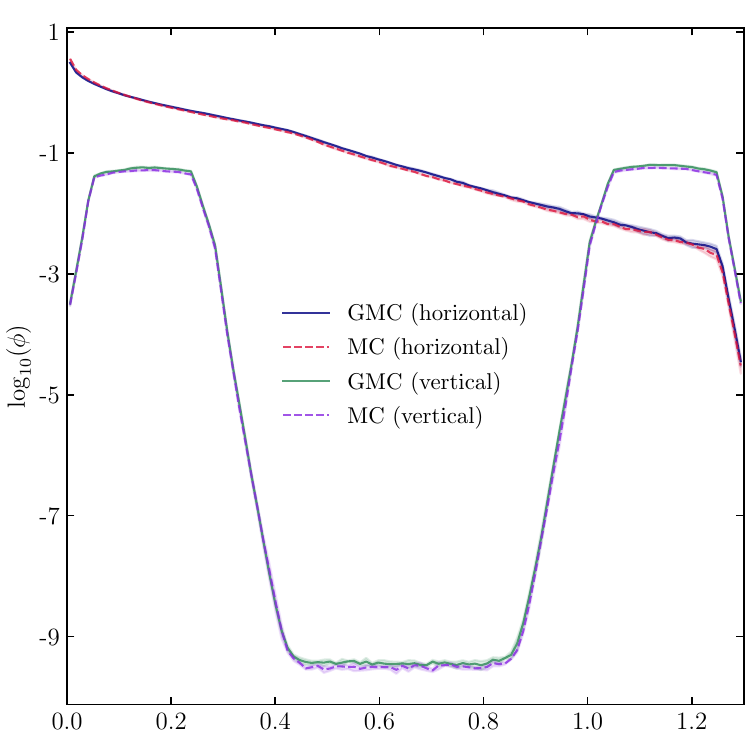}
        \caption{Lineout comparison}
    \end{subfigure}

    \caption{Performance of GMC transport solver on two benchmark problems. \textbf{Top row:} Lattice problem with internal source showing (a) geometric layout on a $7\times7$~cm domain discretized into $112\times112$ cells, (b) scalar flux distribution with MC reference ($10^6$ particles, left) and GMC ($10^6$ particles, right), (c) horizontal and vertical lineout comparisons through the source center. \textbf{Bottom row:} Linearized hohlraum problem with boundary source showing (d) geometric layout with heterogeneous materials on a $1.3\times1.3$~cm domain discretized into $112\times112$ mesh, (e) scalar flux distribution (MC reference with $10^6$ particles, top; GMC with $10^6$ particles, bottom), (f) horizontal and vertical lineout comparisons. The scalar flux is computed via the track-length estimator normalized per source particle.}
    \label{fig:validation}
\end{figure*}

The first benchmark (Figure~\ref{fig:validation}a–c) models a $7\times7$~cm heterogeneous lattice discretized on a $112\times112$ mesh. The background scattering medium has cross sections $\sigma_a = 0$, $\sigma_s = 1$~cm$^{-1}$ (total $\sigma_t = 1$~cm$^{-1}$), while the absorbing regions embedded in a periodic array have $\sigma_a = 9.5$~cm$^{-1}$, $\sigma_s = 0.5$~cm$^{-1}$ ($\sigma_t = 10$~cm$^{-1}$). An isotropic volumetric source is placed in the central $1\times1$~cm subdivision. This configuration tests the \texttt{Internal Model}'s ability to handle volumetric sources and correctly predict flux depression within strongly absorbing media.

Figure~\ref{fig:validation}b presents a split-view heatmap of the scalar flux distribution, with the MC solution ($10^6$ particles) on the left and the GMC prediction ($10^6$ particles) on the right. Qualitatively, the GMC model reproduces all principal solution features, including the characteristic low-flux shadows within the absorbers and the steep gradients radiating from the source region.

The lineout comparisons are shown in Figure~\ref{fig:validation}c, extracted along horizontal and vertical cuts through the mesh. The GMC solutions (solid lines) track the MC reference (dashed lines) closely across the domain, accurately capturing the flux attenuation---which spans orders of magnitude---as particles penetrate the absorbing regions. The agreement in both the scalar flux maps and lineouts is evidence of the \texttt{Internal Model}'s ability to accurately sample from complex internal source distributions in heterogeneous media.

The second benchmark (Figure~\ref{fig:validation}d–f) represents a linearized hohlraum geometry ($1.3\times1.3$~cm) on a $112\times112$ mesh, driven by a 
boundary source at the left wall ($x=0$). The geometry features multiple material regions: black absorbing walls ($\sigma_a = 50$, $\sigma_s = 50$~cm$^{-1}$), red stripes ($\sigma_a = 5$, $\sigma_s = 95$~cm$^{-1}$), green framing regions ($\sigma_a = 10$, $\sigma_s = 90$~cm$^{-1}$), a central blue absorber ($\sigma_a = 95$, $\sigma_s = 5$~cm$^{-1}$), and white near-vacuum regions ($\sigma_a = 0$, $\sigma_s = 5$~cm$^{-1}$). This problem is streaming-dominated in the white regions but features strong wall interactions that test the scattering mechanics. The scalar flux map (Figure~\ref{fig:validation}e) confirms that GMC correctly models the ray-like streaming behavior from the boundary source, as well as the back-scattering from the optically thick walls. The lineouts (Figure~\ref{fig:validation}f) demonstrate quantitative agreement in the flux gradients across material interfaces. 

\subsection*{Convergence and Performance}

Since  GMC  functions as a stochastic surrogate, it is essential to verify that it behaves as a consistent statistical estimator. We conduct a convergence study using total particle counts $N \in \{10^5, 2\times10^5, 4\times10^5, 8\times10^5, 1.6\times10^6, 3.2\times10^6, 6.4\times10^6, 1.28\times10^7\}$. For each particle count, we perform $K=5$ independent simulation runs and compute the cell-averaged standard deviation of the scalar flux,
\begin{equation}
\bar{\sigma}_\phi(N) = \frac{1}{N_{\text{cells}}} \sum_{i=1}^{N_{\text{cells}}} \sqrt{\frac{1}{K-1} \sum_{k=1}^{K} \left( \phi_i^{(k)} - \bar{\phi}_i \right)^2},
\end{equation}
where $\phi_i^{(k)}$ is the scalar flux in cell $i$ for run $k$, and $\bar{\phi}_i = K^{-1}\sum_k \phi_i^{(k)}$ is the ensemble mean. This metric quantifies the statistical uncertainty in the solution as a function of sample size.

\begin{figure*}[htbp]
    \centering
    \begin{subfigure}[b]{0.49\textwidth}
        \centering
        \includegraphics[height=6cm]{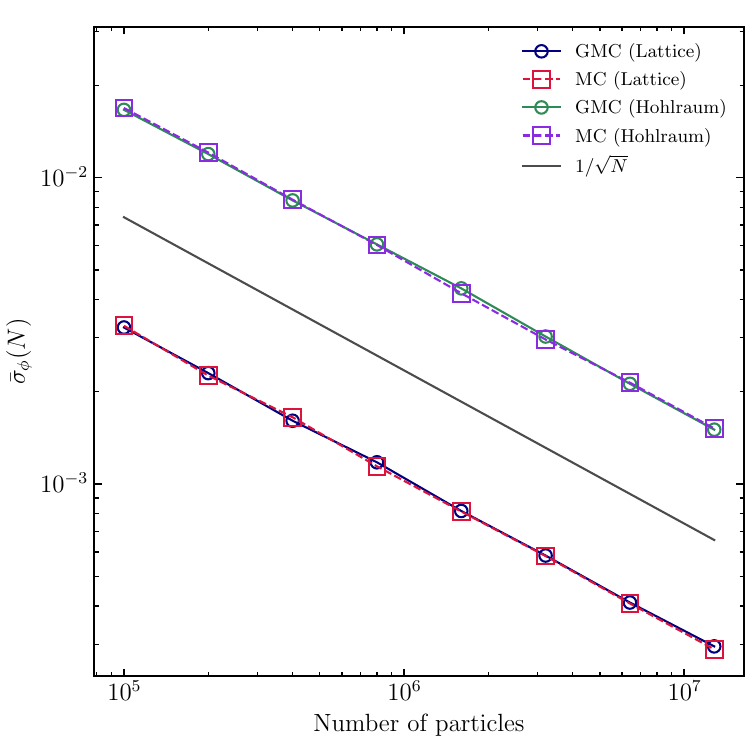}
        \caption{Statistical convergence}
        \label{fig:std_convergence}
    \end{subfigure}
    \hfill
    \begin{subfigure}[b]{0.49\textwidth}
        \centering
        \includegraphics[height=6cm]{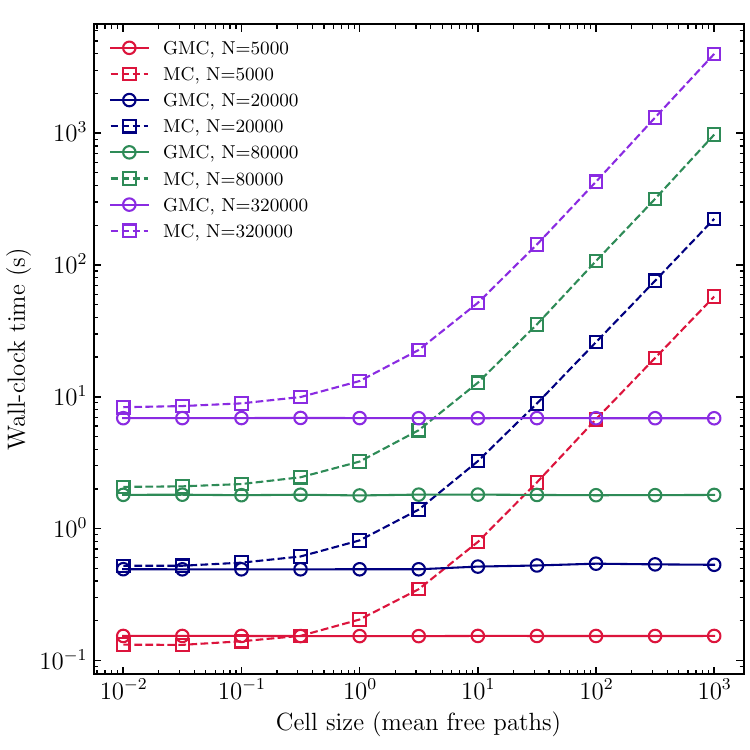}
        \caption{Computational scaling}
        \label{fig:perf_comparison}
    \end{subfigure}

    \caption{%
    Statistical convergence and computational performance of GMC-based transport.
    Figure~\ref{fig:std_convergence} shows the convergence of the cell-averaged standard deviation $\bar{\sigma}_\phi$ in scalar flux estimates as a function of sample size $N$, computed from $K=5$ independent runs at each particle count. Both GMC and standard MC exhibit the characteristic $N^{-1/2}$ convergence rate, confirming that the surrogate preserves the statistical properties of Monte Carlo estimation.
    Figure~\ref{fig:perf_comparison} demonstrates the algorithmic advantage of the surrogate in a pure scattering medium. Standard MC cost (red) increases linearly with the cell's optical thickness $\tilde{W} = W\sigma_s$ (due to the increased number of scattering events required to escape), the GMC surrogate (blue) maintains constant $O(1)$ computational cost per cell transmission via fixed-step ODE integration.}
    \label{fig:converge}
\end{figure*}

Figure~\ref{fig:std_convergence} presents the convergence of $\bar{\sigma}_\phi$ for both the lattice and hohlraum benchmarks. Both the standard MC and GMC methods exhibit the characteristic $N^{-1/2}$ convergence rate expected for Monte Carlo estimators. The GMC curves are coincident with the corresponding MC curves, confirming that the exit samples generated by the flow matching ODE solver are statistically independent and identically distributed (i.i.d.) relative to the learned conditional distribution. This parallel behavior indicates that the intrinsic variance characteristics of the transport process are preserved by the surrogate model without introducing systematic bias or artificial noise correlations.

The primary motivation for the GMC approach is computational efficiency in optically thick regimes. To quantify this, we analyze a purely scattering medium where the computational load is controlled by varying the optical dimensions of the cell. Let $\tilde{W} = W \sigma_s$ and $\tilde{H} = H \sigma_s$ denote the cell width and height in units of mean free paths. Figure~\ref{fig:perf_comparison} compares the normalized runtime of standard MC against GMC as a function of optical thickness.

In standard Monte Carlo, the computational cost scales linearly with the number of scattering events; as the cell becomes optically thicker (larger $\tilde{W}$ and $\tilde{H}$), particles undergo a diffusive random walk involving many collisions before reaching a boundary. This causes the runtime to increase significantly. In contrast, the GMC computational cost is effectively constant, $O(1)$, with respect to the optical thickness. Because GMC integrates a fixed-step ODE (using 12 integration steps with an RK4 solver) to map the entry state directly to the exit state regardless of the underlying number of scatterings, it achieves order-of-magnitude speedups in large, highly scattering cells while maintaining physical fidelity.

\section*{Conclusions}

We have presented Generative Monte Carlo, a paradigm that integrates generative AI directly into the stochastic solution of the Boltzmann equation. By reformulating the cell-transmission problem as a conditional generation task, we demonstrated that machine learning can accurately reproduce the complex, multi-dimensional exit distributions of particles without simulating scattering histories.

This approach maintains the physics fidelity required for scientific applications. Through the use of optical coordinate scaling, a single trained model successfully generalized to diverse heterogeneous geometries, ranging from reactor-like lattices to streaming-dominated hohlraums. The results confirm that GMC preserves the fundamental quantities of transport—specifically the scalar flux and energy deposition profiles—while exhibiting the same $1/\sqrt{N}$ statistical convergence rate as standard Monte Carlo. This consistency ensures that GMC functions not merely as an approximate interpolator, but as a robust statistical estimator compatible with existing transport frameworks.

The most significant advantage of GMC lies in its algorithmic scaling. We showed that while standard Monte Carlo suffers from a linear increase in computational cost as materials become optically thick (the diffusive limit), the cost of the GMC surrogate remains constant ($O(1)$). By effectively teleporting particles across diffusive regions via a fixed-step ODE integration, GMC overcomes one of the longest-standing bottlenecks in radiation transport.

Finally, this framework strategically realigns the field of particle transport with the trajectory of modern computing. As hardware development increasingly prioritizes tensor operations and neural inference acceleration, GMC allows transport codes to natively exploit these advancements. By shifting the computational burden from branching logic and random memory access to contiguous vector operations, GMC paves the way for next-generation simulations that are both physically accurate and computationally scalable on heterogeneous platforms.

Our presentation focused on single-energy, steady results. The extension to energy- and time-dependent calculations  will be presented in future work, but we give the basic ideas here. For time-dependent calculations the particle sampling will be done in space-time to determine when a particle crosses a time step boundary. This requires new training data and presents no further complication. For energy dependence we can readily apply the multigroup method, which breaks the problem into a set of single-energy equations coupled by volumetric sources (a problem equivalent to that solved above). Continuous energy MC could also use our approach by conditioning the model on the cell's composition and having the exit energy be a dependent variable. Unstructured grids present a different, but we believe soluble, problem. In two-dimensions with a triangular grid, for instance, the sampling would be conditioned on the triangle shape (3 numbers at most). Grids with varying element shapes could also be handled by having a trained model for general quads and a separate one for triangles. In addition to these ideas, we believe there are direct applications of the approach to transport in stochastic media and uncertainty quantification.

\acknow{This work is supported by the Center for Exascale Monte-Carlo Neutron Transport, a PSAAP-III project funded by the U.S. Department of Energy, grant number DE-NA003967.}
\showacknow{}
\showmatmethods{} 

\bibliography{pnas-sample}

\end{document}